\documentstyle[emulateapj]{article}

\lefthead{Armitage \& Livio}
\righthead{Stream-disk hydrodynamics}

\begin{document}

\title{Hydrodynamics of the stream-disk impact in
	interacting binaries}

\author{Philip J. Armitage\altaffilmark{1} and Mario Livio}
\affil{Space Telescope Science Institute, 3700 San Martin Drive,
        Baltimore, MD 21218; \\
        armitage@cita.utoronto.ca, mlivio@stsci.edu}     

\altaffiltext{1}{Permanent address: Canadian Institute for Theoretical 
Astrophysics, McLennan Labs, 60 St George St, Toronto, M5S 3H8, Canada}

\begin{abstract}
We use hydrodynamic simulations to provide quantitative estimates
of the effects of the impact of the accretion stream on disks in
interacting binaries. For low accretion rates, efficient radiative
cooling of the hotspot region can occur, and the primary consequence
of the stream impact is stream overflow toward smaller disk radii.
The stream is well described by a ballistic trajectory, but 
larger masses of gas are swept up and overflow at smaller, but
still highly supersonic, velocities. If cooling is inefficient,
overflow still occurs, but there is no coherent stream inward
of the disk rim. Qualitatively, the resulting structure appears 
as a bulge extending downstream along the disk rim. 
We calculate the mass fraction and velocity of the overflowing
component as a function of the important system parameters, 
and discuss the implications of the results for
X-ray observations and doppler tomography of cataclysmic
variables, low-mass X-ray binaries and supersoft X-ray sources.
\end{abstract}	


\section{INTRODUCTION}

Observations of cataclysmic variables, low-mass X-ray binaries,
and supersoft X-ray sources show that accretion disks in these
systems display pronounced deviations from axisymmetry, for
which the most obvious agent is the impact of the gas stream
from the companion star onto the disk (for a review, see e.g.
Livio 1993). The complexity of the
stream-disk interaction mandates the use of hydrodynamic 
simulations to fully explore its consequences.

The dynamics of the accretion stream up to the point of impact
with the disk was described by Lubow \& Shu (1975,
1976). The gas stream leaves the inner Lagrange point in
approximate vertical hydrostatic equilibrium, and flows
inward along a ballistic trajectory. As it does so, inertia 
precludes the stream from adjusting fast enough to maintain 
hydrostatic balance, so that upon reaching the disk edge
the stream is substantially more extended vertically than
the local hydrostatic value. Gas at several scale heights
above the disk midplane may then be able to overflow
the disk rim towards smaller disk radii. Some possible
observational signatures of this were discussed for
X-ray binaries by Frank, King \& Lasota (1987), and
for cataclysmic variables by Lubow (1989).

In this investigation we extend the study of the stream-disk 
interaction by performing
three dimensional calculations of the stream-disk interaction
using the ZEUS hydrodynamics code (Stone \& Norman 1992a,b).
Our goals are to quantify the degree of stream overflow,
to explore the structure of the stream-disk impact region at the
disk rim, and to investigate the efficiency of entrainment
once the gas stream is overflowing the disk. These theoretical
questions correspond closely to observational issues; how
much absorption can be generated by an overflowing stream, how
extended is the hotspot region, and can the stream-disk 
impact generate significant emission at velocities
intermediate between those of the disk and a ballistic gas
stream?

The layout of this paper is as follows. In \S 2 we discuss
and present the results of hydrodynamic simulations of
the stream-disk impact. We compare our results with our previous
calculations which used smooth particle hydrodynamics (Armitage \& 
Livio 1996), and with
two-dimensional calculations (Rozcyczka \& Schwarzenberg-Czerny 
1987). \S 3 quantifies the degree of stream overflow as a
function of the parameters of the stream and disk, while \S 4
analyzes the calculated velocity profile of the stream and
entrained disk gas. \S 5 discusses
how the simulations compare to current observational results,
while \S 6 summarizes our main conclusions.
 
\section{HYDRODYNAMIC SIMULATIONS OF THE INTERACTION}

Hydrodynamic calculations of the interaction of the stream 
with the accretion disk were carried out with the ZEUS-3D
code developed by the Laboratory for Computational Astrophysics
(Clarke, Norman \& Fiedler 1994).
ZEUS-3D is a finite difference code utilizing an artificial 
viscosity to capture shocks, which has been extensively used
and tested for astrophysical fluid dynamics problems. Details
of the algorithms used in the code are described by Stone \& 
Norman (1992a, 1992b).

\subsection{Initial and boundary conditions}

The simulations are performed in Cartesian $(x,y,z)$ geometry,
with cubic cells. The computational box surrounds the impact
point of the stream and the disk, and is aligned with the
$y$ axis in the radial direction and the $z$ axis aligned vertically.
Symmetry about the disk midplane is assumed. The inflow
of the stream and disk are prescribed as boundary conditions,
and the calculations are run until the mass and vertical kinetic 
energy within the box reach an approximate steady state. This
requires $\sim 2$ flow crossing times of the grid. Gravity 
from the accreting star only is included. The grid covers
50 \% of the radius of the disk, and about $45^{\circ}$
in azimuth.

The disk prior to impact with the stream is taken to be
in hydrostatic equilibrium in the vertical direction,
which together with an isothermal vertical structure 
corresponds to a gaussian density profile with scale
height $H_{\rm d}$. At the outer edge of the disk, $R_{\rm out}$,
the sound speed, $c_s$ is set such that the Mach number of the flow
is 30. The disk is always taken to be isothermal in the
vertical direction, for most calculations we additionally
assume that the disk is isothermal radially, so that 
$H_{\rm d} (R) \propto R^{3/2}$. The flaring of the 
disk implied by this means that the strongest interaction
with the overflowing gas will be near the disk rim.

Since the hydrodynamics of the impact are essentially 
those of hypersonic flow past an obstacle, the 
sharpness of the {\em radial} truncation of the disk 
at the outer edge may be expected to be of some importance.
Tidal torques, and possibly the ram pressure of the
stream itself, will tend to produce a sharp edge to
the disk, but this will be opposed by the turbulent
motions giving rise to the viscosity. Consequently, we assume a
gaussian fall-off of the disk midplane density
$\rho_{\rm d0}$ beyond $R_{\rm out}$, with a scale 
length equal to the vertical scale height
$H_{\rm d}$. Interior to $R_{\rm out}$ the midplane
density is constant with radius.

The initial conditions for the gas stream are similar
to the results of the integration of the stream
dynamics described by Lubow (1989). We assume 
that the stream density falls off with
distance from the stream center as a gaussian,
with scale length $H_{\rm s}$, and central density
$\rho_{\rm s0}$. The initial vertical velocity
is taken as $v_z \propto -z$, with the constant of
proportionality equal to $0.6 c_s$, implying modest convergence
of streamlines towards the disk midplane. The Mach
number of the stream at the disk edge is 30, and
the angle between the stream and the radial direction
at the impact point is $15^{\circ}$. The stream gas 
is taken to have the same temperature as the outer 
edge of the disk, for many systems this will be a 
reasonable assumption. In cases where the outer parts 
of the disk are much hotter than the surface layers 
of the secondary, ballistic stream overflow would of
course be inhibited. 

Slow inflow through a viscous disk implies that 
$\rho_{\rm d0} \gg \rho_{\rm s0}$. To within factors
of order unity, a steady-state disk described by the 
Shakura-Sunyaev (1973) $\alpha$-prescription 
for the viscosity has,
\begin{equation}
 { {\rho_{\rm s0} \over {\rho_{\rm d0}} }} \simeq
 6 \alpha \left( c_s \over v_s \right) \left( H_{\rm d}
 \over H_{\rm s} \right)^2,
\label{eq_rho_ratio}
\end{equation}
with $v_s$ the stream velocity at the impact point. 
For the parameters of our calculations, and assuming 
a typical $\alpha$ value of 0.1 for interacting binary
disks, $ \rho_{\rm s0} /  \rho_{\rm d0} \sim 10^{-2}$,
and we use this density ratio in the simulations.

Reflecting boundary conditions are imposed at the disk mid-plane,
$z=0$. Inflow boundary conditions are specified where the disk
flows into the computational volume, at $x=0$, and at $y=0$
over the area where the stream enters. Over the remainder
of the $y=0$ plane, and for the other three faces of the 
computational box, outflow boundary conditions are used.
Additionally, for the adiabatic simulations, we allow 
outflow through the $x=0$ face where the calculated inflow 
density for the disk is below $10^{-12}$ of the central disk 
value (i.e. at high $z$). This is required in order to prevent 
artefacts arising near the boundary at several scale heights
above the disk midplane.

For reference, the calculations use second order (van Leer 1977)
interpolation for all advected quantities, a dimensionless
coefficient of artificial viscosity $C_2 = 2.0$, and a Courant
number of 0.5. 

\subsection{Cooling}

We have run simulations with three equations of state; isothermal,
adiabatic (with no cooling), and one with optically thin
radiative cooling included. The isothermal and adiabatic with no
cooling cases bracket the range of possible behavior, the calculation
including cooling is more realistic although there are many complexities
that we do not attempt to model in these calculations.

If cooling is so rapid that the equation of state
is effectively isothermal, then the main parameters of the
problem as set up here are the Mach number of the flow, the ratio 
of disk to stream density, and the ratio of disk to stream scale 
heights. The results in this limit are independent of the absolute
disk density, or equivalently of the mass transfer rate in the
stream. This degeneracy is broken when more realistic cooling is 
included, dense gas then cools faster than diffuse material, and 
the results depend on the actual value of $\dot M$.

For a simple description (following Blondin, Richards \& Malinowski 1995)
the cooling time is given approximately by,
\begin{equation}
 t_c \approx { {k_B T_{\rm gas}} \over {n \Lambda (T_{\rm gas})} },
\label{t_cool}
\end{equation}
where $k_B$ is the Boltzmann constant, $n$ is the number density of
electrons, and $\Lambda$ is the cooling rate in ergs cm$^{-3}$ s$^{-1}$.
We use for $\Lambda (T_{\rm gas})$ the cooling function given by 
Dalgarno \& McCray (1972, see also Cox \& Daltabuit 1971), for
which $t_c$ (defined here as the time required to cool to 
$T_{\rm gas} / 2$) is given at a fiducial temperature of 
$10^6$ K by,
\begin{equation}
 t_c \simeq 10^2 \left( { n \over {10^{10} {\rm cm}^{-3}} } \right)^{-1} \ {\rm s}.
\label{t_cool_numeric}
\end{equation}
The qualitative hydrodynamics of the flow will then be determined
(Blondin, Richards \& Malinowski 1995) by the ratio of $t_c$ to the
dynamical time at the edge of the disk, $t_d \sim \Omega^{-1}$,
where $\Omega$ is the Keplerian angular velocity. For many systems,
$t_d$ will be of order $10^3$ s or longer.

For an estimate, we assume that the emission from the hotspot arises
from material at comparable density to that in the stream (the
post shock density will be higher than this, so cooling will be
{\em more} rapid in the optically thin limit). Then to order
of magnitude,
\begin{eqnarray}
 n \sim 2 \times 10^{12} \ \left( \dot{M} \over {10^{-10} M_\odot {\rm yr}^{-1}} \right)
 \left( H_{\rm s} \over {5 \times 10^9 {\rm cm}} \right)^{-2} \\ \nonumber
 \times 
 \left( v_{\rm s} \over {300 {\rm kms}^{-1}} \right)^{-1} \ {\rm cm}^{-3}.
\label{stream_density}
\end{eqnarray}
Comparison of
this equation with equation (\ref{t_cool_numeric}) indicates that
in the regime where optically thin cooling described by a cooling
function is a valid approximation, typical accretion rates imply 
a cooling time $t_c \ll t_d$. Thus, at low accretion rates 
we expect the gross hydrodynamics
of the stream-disk impact to be well described by an isothermal
equation of state. The flow may be expected to depart from isothermal
behavior in the low density regions several scale heights from the
midplane, and to explore this regime we have run several simulations
including optically thin cooling. 

\subsection{Dependence of the effective equation of state on
        the accretion rate}
        
The above argument suggests that the accretion rate will generally
be high enough that cooling will be rapid, {\em provided} that the 
emission is optically thin. In this regime an isothermal equation of
state should provide a good description of at least the higher density 
regions of the flow, within a few scale heights of the disk midplane. At 
higher accretion rates the assumption that the optical depth to the
emitting regions of the hotspot is small will break down, and qualitatively
we expect that the hydrodynamics will be better approximated by an
adiabatic equation of state. As we demonstrate later, the flow in
these two regimes is very different -- here we estimate where the
transition between them occurs.

The radiation emitted as the shocked stream gas cools must escape,
at a minimum, through the hot ($\sim 10^6$ K) shocked layer and the 
relatively cool ($\sim 10^4$ K) inflowing stream material. We consider
first the opacity due to the cooler stream.
The temperature of the stream at the outer edge of the disk will
be comparable to the surface temperature of the mass losing star,
with a modest enhancement from compressional heating as the stream
converges on the disk midplane (in some systems irradiation of the 
stream by the disk or hotspot might also be important -- this would be
particularly significant if the stream gas was able to be highly ionized 
prior to striking the disk). The dominant opacity source is therefore
likely to be H$^{-}$ scattering. An analytic approximation to the
opacity in this temperature range is given by Bell \& Lin (1994), in
units of cm$^{2}$ g$^{-1}$,
\begin{equation}
 \kappa = \kappa_0 \rho^{1/3} T^{10}
\end{equation}
with $\kappa_0 = 10^{-36}$. The column density for escape of the 
hotspot radiation will evidently be $\sim \rho_{\rm s0} H_{\rm s}$,
for which the optical depth is,
\begin{equation}
 \tau \sim {{\kappa \dot{M}} \over {\pi H_{\rm s} v_{\rm s}} }.
\end{equation}        
Imposing the condition that $\tau=1$ then gives an estimate for the
critical accretion rate for the transition between isothermal and
adiabatic behavior. Using the numerical values adopted previously 
we obtain,
\begin{eqnarray}
 \dot{M}_{\rm crit} \sim 2 \times 10^{-9} 
 \left( {H_{\rm s} \over {5 \times 10^{9} {\rm cm}}} \right)^{5/4}
 \left( {v_{\rm s} \over {300 {\rm kms}^{-1}}} \right) \\ \nonumber 
 \times 
 \left( {T \over {10^4 {\rm K}}} \right)^{-7.5} 
 M_\odot {\rm yr}^{-1}.
\end{eqnarray}
For accretion rates below this value cooling should be efficient,
above it the radiation will be trapped by the inflowing stream gas 
and the initial cooling of the hot spot region will occur via
adiabatic expansion. 

The opacity of the hot shocked layer can also limit the applicability 
of the efficient cooling assumption. At $\dot{M}_{\rm crit}$, the 
central stream density for the parameters used previously is 
$\sim 5 \times 10^{-11} \ {\rm gcm}^{-3}$. While the shock is
still isothermal, the compression could be as large as the Mach number
($\sim 30$) squared,
giving a shocked layer with a density of $\sim 5 \times 10^{-8} \ 
{\rm gcm}^{-3}$. For an opacity due to electron scattering, $\kappa = 0.348$,
and $\tau=1$ requires a thickness of the hot layer of order $10^8 \ {\rm cm}$.
As this is indeed roughly the thickness expected of the shocked region,
it suggests that the critical accretion rate for the emission to 
become optically thick from the opacity of the shocked layer itself 
is comparable to, or somewhat less than, the critical value estimated
above. We conclude that, with considerable uncertainties, both 
arguments give $\dot{M}_{\rm crit} \sim 10^{-9} \ M_\odot {\rm yr}^{-1}$.  

Clearly the above analysis is extremely crude, and no substitute for 
a proper radiative transfer calculation of the hotspot region. {\em A priori} 
it would definitely be wise to keep an open mind as to whether a 
system with a given accretion rate would be better described theoretically 
with isothermal or adiabatic models. However it suggests that the 
hotspot region in low accretion rate cataclysmic variables might well 
be able to cool efficiently, whereas nova-like variables and 
supersoft X-ray sources with accretion rates much larger than
$\dot{M}_{\rm crit}$ are almost certainly unable to do so. 

In the following section we present models for both limiting cases. 
For low accretion rate systems we use an equation of state that is
either isothermal, or which includes radiative cooling in the 
optically thin limit. These are qualitatively very similar, and
differ mainly in the very low density regions where optically thin
cooling becomes ineffectual. For models intended to represent better
higher accretion rate systems we use an adiabatic equation of
state without any cooling. Real systems, where cooling {\em will} occur
in the low density parts of the flow, would be expected to have
intermediate properties between the limiting extremes presented
here.

\subsection{Structure of the impact region for isothermal and 
	    adiabatic equations of state}

Figure 1 (Plate XXX) illustrates the primary differences between 
the simulations using isothermal (or efficient radiative cooling) 
and adiabatic equations of state. The left panels show isodensity 
surfaces for an isothermal calculation, plotted at a density that 
is $10^{-3}$ of the central disk density, or $1/10$ of the central 
stream density. The colors represent the radial velocity on this 
surface. 

For this equation of state, the uppermost part of the stream where the 
density significantly exceeds that of the disk is able to flow smoothly over 
the rim of the disk in the manner envisaged by Lubow \& Shu (1976), and
Frank, King \& Lasota (1987). We do not observe any instabilities 
that lead to rapid entrainment of the stream as it flows over the disk 
surface, though the leading edge of the stream is deflected by the 
ram pressure of the disk gas. More interestingly, there is a fairly 
broad region of the disk, downstream of the region where the stream 
is overflowing, which possesses significant inward velocity. This 
is material that has been brought to an intermediate velocity between 
that of the stream and the disk, as a result of the stream-disk 
interaction. In our simulations including radiative cooling this 
gas is hot, and thus a potential source of emission with a radial 
velocity that is highly supersonic, but less than that of the 
ballistic stream.

The right panels of the Figure show the results for an adiabatic 
equation of state. Here, the isodensity surface is plotted at a
density that is $10^{-2.75}$ of the central disk density, and the 
colors represent $\log (c_s)$ on that surface. The details of 
this calculation will be discussed in Section 2.6, but it is 
obvious from the Figure that in this case the upper regions of 
the stream are {\em not} able to flow unimpeded over the surface of 
the disk. Rather, the qualitative impression is that the impact 
leads to a hot bulge extending along the rim of the disk. This 
material is not in hydrostatic equilibrium, and expands outwards 
downstream of the impact point. Material is still overflowing 
towards smaller radii, but it is not in the form of a coherent 
stream. We also note that even in the absence of radiative 
cooling, the temperature on the plotted surface declines 
rapidly (on the scale of a few $H_{\rm s}$) moving downstream 
of the impact point.	    	    

\subsection{Results for an isothermal equation of state /
	efficient radiative cooling}

Figures 2 through 4 show in greater detail the results of the
calculation that was illustrated in the left panels of Fig.~1. 
This employed an isothermal equation of state, and a ratio of stream
to disk scale heights $H_{\rm s} / H_{\rm d} = 2$. The
computational volume extended vertically to 5 disk 
scale heights, and in the $x$ and $y$ directions for
16 disk scale heights. We discuss later the differences
in simulations including efficient radiative cooling, but the
general behavior seen in those runs is the same as in
the isothermal case. The computational grid used was
$176 \times 176 \times 55$ cells. We have run simulations 
including radiative cooling at resolutions of up 
to $192^2 \times 60$, and adiabatic calculations at
up to $128^2 \times 40$. We note, however, that there 
is a sharp vertical density gradient in the disk in the
region, several scale heights above the midplane, where 
the stream is overflowing. Higher resolution might thus 
reveal additional interactions that are not resolved in
these calculations.  

Fig.~2 shows density contours and velocity vectors in the $x-y$ plane, 
with the stream entering from the bottom in each panel, and the
disk gas flowing in from the left. In the
disk midplane (top left panel), the stream is stopped rapidly at the
edge of the disk by the much denser disk material. There
is no `splashing' or ricocheting material. Moving upwards,
at around 2 disk scale heights above the midplane (top right panel) the 
densities of the stream and disk start to become comparable.
Here we resolve the twin shocks in the disk and stream gas
seen in earlier two dimensional calculations (Rozcyczka \& 
Schwarzenberg-Czerny 1987) and predicted analytically
(e.g. Livio 1993). We also see that the velocity field 
of the disk gas downstream of the impact point has been 
disturbed by the impact -- this material now has a significant
radial velocity inward. At higher elevations the stream 
rapidly begins to dominate. At $z=2.5 H_{\rm d}$ the
stream is already able to penetrate to the inner boundary of the
computation (at $\sim 0.5 R_{\rm out}$), and at $z=3 H_{\rm d}$ 
the stream is overflowing almost freely, with some modest
deflection of the edge facing the oncoming disk material.

Comparing these results with the two dimensional calculations
of Rozcyczka \& Schwarzenberg-Czerny (1987), which have
close to identical resolution in the plane, we see the
same shock structure and some evidence of unstable flow
in the mixed stream and disk layer. However our 3D simulations
do not support the idea of a hotspot region that is highly extended
along the disk rim. In our calculations shocks, and hence
emission (given the implicit assumption of rapid cooling),
are restricted in the midplane to the part of the disk
directly struck by the stream. At higher $z$, the shocks
are extended, but primarily along the radial direction.
The two dimensional nature of earlier calculations, 
together with the assumption of a slow fall-off of 
the density at the disk edge, are likely to be responsible 
for these differences.  

Fig.~3 shows density contours in a vertical slice taken along
the initial stream flow direction. Although the stream is
converging towards the disk midplane at all radii, the
interaction between the stream and the disk is concentrated
at the disk rim. Inward of the rim, there is a clear 
density minimum between the upper surface of the disk and
the lower part of the overflowing stream. This is in accord
with the prediction of Lubow (1989), and it suggests that the
final fate of the stream, not modeled here, will be to
reimpact the disk near to the point of the ballistic stream's
closest approach to the central object. We note that the
separation between overflowing stream and disk seen in 
this simulation is accentuated by the strong flaring
of the disk scale height implied by the isothermal 
assumption, $H_{\rm d} \propto R^{3/2}$. However,
weak interaction between stream and disk interior to
the rim is also seen in a simulation where we took
$H_{\rm d} \propto R^{9/8}$, appropriate if the
disk central temperature scales with the run of
effective temperature $T_{\rm eff} \propto R^{-3/4}$
(a `standard' disk). 
Of course stronger interaction {\em would} be expected
if the disk atmosphere was much more vertically extended
than a gaussian.

Fig.~4 shows the density on vertical cylindrical surfaces
at constant radii in the disk (i.e. the co-ordinates are
$R \phi$,$z$). Three radii are depicted, at $R > R_{\rm out}$,
showing just the cross-section of the stream, at the disk
rim where interaction is strongest, and where the stream
is overflowing the disk interior to the rim. The distortion
of the stream by the ram pressure of the disk can be seen,
and it is clear that the main effect of the stream on the
disk rim downstream of the impact point is to greatly
reduce the density at high $z$. The stream acts as 
a windbreak in the flow, and truncates the disk
in the vertical direction. For these parameters
a very small quantity of disk gas is deflected 
{\em over} the stream -- this phenomenon is stronger 
in simulations where radiative cooling is included 
and can lead to a hot wake of material downstream
of the impact point along the rim.

These results are obtained for strictly isothermal flows.
However, the flow in simulations including
explicit radiative cooling is extremely similar in the
regime where the cooling is much more rapid than the
dynamical timescale. As argued earlier, this should
be applicable to disks that are optically
thin at all reasonable accretion rates. 

Fig.~5 compares the results of the isothermal calculation
with two simulations which include cooling. The cooling simulations
assume central disk electron densities of $10^{14}$ and $10^{13}$
particles per cm$^{-3}$, which correspond to accretion
rates of order $10^{-10}$ to $10^{-11} \ M_\odot {\rm yr}^{-1}$.
The cooling simulations used a $128^2 \times 40$ grid.
A floor temperature of $10^4$ K was imposed.
We plot vertical profiles of density and, for the runs
with cooling, temperature, at a point 20$^\circ$ downstream
of the stream impact point just interior to the disk rim.

The density profiles show that the low density wake behind
the overflowing stream persists in all three runs. In the
runs incorporating cooling the density at high $z$ can be
two to three orders of magnitude higher than for the isothermal
case -- this is a consequence of the overspill of disk gas
{\em above} the stream mentioned earlier. The absolute density
though is very low in all cases. High temperatures of $10^5$ K
and above are seen in a layer above the disk surface, this
is material that has been shock heated by the interaction with 
the stream and which has passed beneath the overflowing gas.
This feature is not adequately resolved in these
simulations.

\subsection{Results for inefficient cooling}

Fig.~6 summarizes the results for a simulation with an
adiabatic equation of state and no radiative cooling. 
The grid was again $128^2 \times 40$ cells, and
$H_{\rm s} / H_{\rm d} = 2$. The remaining initial 
conditions were as for the isothermal and radiative
cooling runs, except that the stream injection point
was moved slightly further from the $x=0$ boundary 
in order to reduce effects arising from the finite 
volume of the computational box. 
 
Qualitatively, the distinction from the isothermal
calculations is that in this simulation the hot
shock-heated gas expands in all directions and
disrupts the coherent overflowing stream seen 
previously. In the disk midplane, `splashing'
of hot material downstream of the
impact point is evident, though the velocities 
of this gas are substantially less than those 
of the disk. At $z=3 H_{\rm d}$
there is overflow in a broad fan over the disk
surface. Most of this material appears to be 
stream gas that has suffered a strong deflection 
upon reaching the disk edge, though an examination 
of the velocity field also shows disk material 
at high $z$ that is being deflected inwards as
a result of the stream interaction.

A slice along the initial stream flow
direction now resolves clearly the shocks in
the stream and disk gas. The overflowing 
component has a velocity that is predominantly 
in the $x-y$ plane, with a small (a few times $c_s$) 
vertical component. Further from the disk midplane significant 
vertical velocity away from the disk {\em is} seen,
suggesting that with this equation of state a 
substantially greater absorption column will be 
generated for lines-of-sight well away from the 
disk plane.  As shown
later, the actual fraction of the stream 
overflowing the disk rim is comparable to that in the
isothermal case, but the structure of the flow
is here very distinct.

\subsection{Global effects}

The ZEUS simulations discussed in this paper suggest a
picture in which efficient radiative cooling leads to 
a coherent stream overflowing the disk in a manner 
very similar to that described by Lubow (1989). Conversely,
if cooling is inefficient, there is a `splashing' of
stream material off the disk edge. Similar quantities
of material can flow inward in this regime, but there is
no coherent stream and gas is also thrown outwards and
upwards. 

In order to examine the qualitative differences in global
effects between the isothermal and inefficient cooling cases,
we performed two three-dimensional smooth particle hydrodynamics
(SPH) calculations. The results are shown in Fig.~7.
These calculations have vastly
lower resolution of the impact region than the ZEUS simulations,
but cover the whole disk and have been evolved for long enough
($\sim 10$ binary orbits) to allow the disk to relax in the
binary potential. We show an isothermal calculation (similar to that
described in Armitage \& Livio 1996), and a calculation with
a polytropic equation of state ($\gamma = 1.1$) and system
parameters appropriate to the supersoft source CAL87 (Callanan \&
Charles 1989; Gould 1995).

Similar features are evident in the SPH calculations as in
the finite difference simulations. The isothermal run shows
a narrow overflowing stream, while the polytropic calculation
leads to a much more vertically extended spray of material
and a bulge of outflowing gas near the impact point. Negligible
masses of material escape the accreting star's Roche lobe
in either simulation. Although the choice of a value for $\gamma$ in
such a calculation is essentially arbitrary, it is clear from 
Figure 7 that the trend is such that systems where cooling is
inefficient should display absorption at much higher elevations
above the disk plane than those well described by an isothermal
equation of state.

\section{DEGREE OF STREAM OVERFLOW}

To quantify the degree of stream overflow in the simulations, 
we compute the integrated radial mass flux through vertical $x-z$ slices
which has a radial velocity $v_R \ge v_{\rm cut}$. For $v_{\rm cut} <
22 c_s$ -- the initial inflow velocity of the stream gas at the
grid boundary -- this
quantity is conserved in the absence of interaction with the disk. 
Fig.~8 shows the radial mass flux as a function of $R / R_{\rm out}$
(where the $R$ for each $x-z$ slice is evaluated along the line 
of the initial stream flow direction), for a variety of threshold 
radial velocities.

Outside the outer edge of the disk, all the curves 
coincide and are flat, reflecting steady-state undisturbed 
stream flow. For the runs with an isothermal or cooling 
equation of state, the amount of material flowing inward 
at the fastest velocity, $\ge 20 c_s$, drops sharply as the disk
rim is reached, and the central portions of the gas stream are 
stopped by the denser disk gas. In this case, and as noted
previously, the main interaction occurs close to the disk rim,
after which the highest velocity curve remains fairly flat
at a lower level than initially. This mass flux is comprised
of material that has overflowed the disk at large enough $z$
to avoid strong interaction with the disk. For the illustrated
case with $H_{\rm s} / H_{\rm d} = 2$, this is around 5\% of
the stream mass transfer rate.

From the Figure, it is evident that a significantly larger
quantity of mass flows inward at velocities that are smaller
than the ballistic stream velocity, but still highly supersonic. 
This is either stream material slowed by strong interaction with
the disk, or disk gas entrained by the overflowing stream. For
the case shown in Fig.~8, a radial mass flux of around 20\%
of the initial stream mass transfer rate is flowing inward
with $v_R \ge 5 c_s$ at the innermost boundary of the simulation
volume, at $\sim 0.6 \ R_{\rm out}$. We analyze the observational
signatures of this material more fully in the following Section. 
We also note that close to the rim, the total mass of gas flowing
inward supersonically exceeds the stream mass flux by a factor
of $\sim 2-3$, though this does not extend far inward.

For the calculations with an adiabatic equation of state, and no
cooling, the curves shown in Fig.~8 are different. In particular,
there is a prominent dip in the curve with the highest velocity
cut near the outer rim of the disk, corresponding to the position
of the shocks seen in Fig.~6. However, although the overflowing
gas forms much less of a coherent stream in this case as compared 
to the isothermal simulations, the total mass moving inward
is comparable. As with the calculation including cooling, 
around 10\% of the stream mass flux overflows the disk at
approximately ballistic velocities (although the stream
is anything but ballistic in this limit), with 
substantially more gas flowing inward at lesser velocities.

For an estimate of the fraction of the stream mass
transfer rate that overflows the disk, we assume that 
{\em all} the mass in the stream above some critical 
$z = z_{\rm crit}$ manages to overflow the disk, while the 
stream below that height is stopped by the collision with 
the disk. For a simple approach, we take $z_{\rm crit}$ 
to scale with the height where the stream density equals
that of the disk (e.g. Livio, Soker \& Dgani 1986),
\begin{equation}
 z_{\rm crit} = \beta { {H_{\rm d} H_{\rm s} } \over 
 \left( H_{\rm s}^2 - H_{\rm d}^2 \right)^{1/2} }
 \left( \ln { \rho_{\rm d0} \over \rho_{\rm s0} } \right)^{1/2},
\label{fitting_formula}
\end{equation}
where we have included a free parameter $\beta$ as a scaling
of this simple estimate.

Fig.~9 shows the results of 3 simulations, all 
incorporating radiative cooling, with varying
ratios of stream to disk scale height. Points are plotted
for three radial velocity cuts, together with a curve showing the
fitting formula given in equation (\ref{fitting_formula}). 
As expected, the overflowing mass fraction is a strong function of
$H_{\rm s} / H_{\rm d}$. For these parameters, $H_{\rm s} / H_{\rm d} = 1.5$
implies that only $\sim 2 \times 10^{-4}$ of the stream passes
inward of the disk rim without strong interaction with the disk,
although almost two orders of magnitude more material is 
moving inward at velocities greater than 5 times the sound speed. 
The naive fitting formula, with an appropriate choice of
$\beta$ (which is however close to unity), provides a
good approximation to the simulation results.

\section{VELOCITY MIXING BETWEEN STREAM AND DISK GAS}

To quantify the observable effects of the stream-disk 
interaction on the velocity of the overflowing stream,
we have computed the expected distribution of radial velocities,
$v_R$, of gas along lines of sight to the central accreting
object. We define the elevation angle of the line of sight
to the disk midplane as $\theta$, and consider viewing the
system at a phase $18^\circ$ downstream of the impact point 
(i.e along a diagonal line in Fig.~2 extending from the lower
right corner towards the central object which is out of the frame 
beyond the upper left corner). We compare the results with those
expected from a freely flowing stream in the absence of the 
interaction with a disk, obtained by running a simulation
with the same stream initial conditions but no disk. This
comparison calculation displays close to ballistic flow,
but does include stream pressure effects. All the calculations 
in this Section used $H_{\rm s} / H_{\rm d} = 2.5$, so that a 
significant fraction of the stream overflows the disk rim.

Fig.~10a shows the results for an angle of elevation above the
disk midplane of $\theta = 12^\circ$. This line of sight in
the isothermal or cooling simulations probes freely overflowing 
gas at high $z$ that has not undergone strong interaction
with the disk. Consequently, we find that the radial velocity
distribution is almost identical to that which would arise 
from the same ballistic stream if the disk were absent. The
distribution has a width of $\sim 5 c_s$, with a sharp
cut-off at large negative $v_R$.

Fig.~10b shows the results closer to the disk midplane, at
$\theta = 9^\circ$. This line of sight probes gas which
has had significant interaction with the disk. The
interaction leads to reduced (less negative) radial
velocities in the column of gas, corresponding to 
the ram pressure induced bending of the inner edge of
the stream seen in Fig.~1, and a narrower distribution
of $v_R$. The deviations from the ballistic prediction
for the line of sight radial velocity are here at the 
$\sim$ 10\% level.

In general we find that for
these relatively large angles above the disk midplane,
the column of gas at large negative $v_R$ greatly exceeds that of the
undisturbed disk at $v_R \approx 0$. These velocity
differences with the ballistic expectation are therefore 
potentially observable as absorption against the central
object. For lines of sight that are closer to the disk midplane,
we find gas with a range of radial velocities that become
steadily closer to $v_R = 0$ as $\theta$ decreases. However, 
there is always a substantial component {\em at} $v_R = 0$ 
which arises from the undisturbed disk interior to the
overflowing stream. This much lower velocity gas would therefore
only be observable in emission.

Finally, we show in Fig.~10c the result for an adiabatic 
equation of state. As expected from the flow pattern
seen in Fig.~6, the adiabatic equation of state 
leads both to a greater column of absorbing gas 
at high angles above the disk midplane, and a
much larger spread of velocities of that gas. 
As compared to the isothermal simulations, the
mean velocity of the gas along the line of
sight is approximately halved, and there is 
a very broad spread of velocities extending even
to positive $v_R$.

\section{COMPARISON WITH OBSERVATIONS}

Three classes of observations potentially probe the
interactions modeled here: X-ray absorption `dips'
in nearly edge-on systems, which may track the column
of overflowing gas; observations of the hotspot
region where the shocks occur; and doppler tomography
analyses which may be able to distinguish the
velocities predicted to arise as a consequence of the
stream impact. We briefly discuss these in turn.

Dips are seen in the UV and X-ray light curves of 
nearly edge-on low-mass X-ray binaries (LMXBs -- e.g
White, Nagase \& Parmar 1995),
intermediate polars, and dwarf novae (Long et al. 1996;
Szkody et al. 1996). In the case
of LMXBs, these have been interpreted as arising from
the stream-disk interaction, either in the form
of an overflowing component (Frank, King \& Lasota 1987),
or as a bulge excited on the rim of the disk (Hellier \& Mason 1989; Mason 1989).
Similar features seen in CVs also seem to be consistent
with this interpretation. In particular, in Z Cha the
dips in a UV lightcurve were seen to become shallower
during the course of a superoutburst, which would
be consistent with a declining mass transfer rate
through the event (Harlaftis et al. 1992). For the
dips observed in the intermediate polars FO Aqr, EX Hya,
BG CMi and AO Psc, and in 
the dwarf nova U Gem in X-ray light curves (Hellier,
Garlick \& Mason 1993), the dips are observed to become shallower 
with decreasing orbital period of the systems. This
is consistent with a statistically smaller mass
transfer rate in shorter period systems (e.g. Warner 1987;
Patterson 1984), and an
interpretation of the dips as arising from the
stream-disk impact. It is also possibly fortuitous, 
as both the simulations and common sense suggest that 
the absorbing column should be a strong function of 
the system inclination. Moreover, these observations do
not discriminate between the overflow and bulge 
models for the absorbing column, and there remains
the observation of dips at a plethora of
phases other that $\sim 0.8$ (e.g. Mason 1989; Thorstensen et al. 1991), 
where the impact occurs.
Some of these may arise at the point where the stream
converges and reimpacts the disk, which occurs at a binary phase 
of $\sim 0.6$ for a variety of mass ratios (Lubow 1989),
or be generated due to an eccentric
disk in low mass ratio systems. The latter possibility
is suggested by the observations of Harlaftis et al.,
and supported by theoretical calculations (Whitehurst 1988a,
1988b; Lubow 1991, Armitage \& Livio 1996; Murray 1996).

We also note that models for the structure of the accretion disk
rim in supersoft X-ray sources require significant `splashing'
at the stream-disk impact point (Meyer-Hofmeister, Schandl \& 
Meyer 1997). Since the accretion rates in these systems are
extremely high ($\sim 10^{-7} \ M_\odot {\rm yr}^{-1}$, e.g. 
van den Heuvel et al. 1992), this requirement is
consistent with the results presented here. The effects of
strong irradiation from the accreting object are an 
additional important complication in both these systems and
in low-mass X-ray binaries (see, e.g the simulations
reported by Blondin 1997).

The simulations reported in this paper suggest that
emission from the pair of shocks generated by the
impact of the stream with the disk occurs over a 
smaller azimuthal extent than was implied by earlier
calculations restricted to two dimensions. This is
in general agreement with observations, for example
of Z Cha, where Wood et al. (1986) obtain a hotspot
extent of $\sim 4^\circ$ in azimuth, and IP Peg, where
Wood \& Crawford (1986) find an extent of
$\sim 3-7^\circ$. The radially extended shock 
structure seen in the simulations is consistent
with an apparently greater azimuthal extent of
the hotspot when seen in ingress as compared to 
at egress, this is seen in IP Peg (Wood \& Crawford 1986).

Optical spectroscopy provides a further probe of the
absorption and emission expected to arise from an
overflowing stream. For example, Hellier (1996) 
reports observations of the nova-like variable V1315 Aql.
In this system, absorption that has been attributed 
to stream overflow is seen at phases close to the point
where the stream impacts the disk. Further along the stream
trajectory, emission is seen at velocities lower than 
the free-fall prediction. This would appear to be 
consistent with an approximately ballistic overflow 
model. However doppler mapping of another nova-like 
variable, SW Sex, shows emission that, if attributed to 
the stream-disk impact, seems more consistent with 
a bulge along the disk rim (Dhillon, Marsh \& Jones 1997). 
This is also the interpretation advanced for doppler maps 
of the dwarf nova WZ Sge (Spruit \& Rutten 1997), where the 
accretion rate is much lower.

Clearly, if the interpretation of all of these observations 
is correct, they do not fit nicely into the simple picture 
presented earlier, where the accretion rate determines 
the geometry of the stream overflow. However
it is at least encouraging that observations, for example of OY Car 
(Billington et al. 1996; see also Bruch, Beele \& Baptista 1996), 
may be capable of constraining the radial location of 
absorption features. This, together with theoretical predictions of 
the velocity fields arising from the stream-disk interaction, 
should help to disentangle the structure of the emission and 
absorption regions generated by the impact.

\section{SUMMARY AND CONCLUSIONS}

In this paper, we have discussed high resolution simulations
of the stream-disk interaction in binary systems. We find
as our main result that the qualitative behavior of the
flow depends on the efficacy of cooling in the shock 
heated gas created by the impact. If cooling is efficient,
specifically if the cooling time is much shorter
than the dynamical timescale $\Omega^{-1}$ in the disk,
then the upper reaches of the stream freely overflow
the disk. This approximately ballistic flow is as
described by earlier calculations (e.g. Frank, King \& 
Lasota 1987; Lubow 1989), and is expected to
terminate in a secondary hotspot where the stream
reimpacts the disk close to its closest approach
to the accreting object. We see no evidence for
hydrodynamic instabilities leading to rapid entrainment 
of the entire stream as it flows over the disk.
Additionally we find
significantly larger quantities of mass that undergo
strong interaction with the disk (or, are entrained
disk gas), and flow inward with intermediate velocities.
This gas may dominate the observed apparent velocity of the
overflowing material at intermediate angles of disk
inclination, and is likely to be visible in emission.
We have argued that this efficient
cooling mode of stream overflow is appropriate to
CVs at low accretion rates, where the
hotspot emission is emanating from regions that
are not optically thick.

Conversely, if cooling is inefficient, then the interaction
leads to a flow better envisaged as an explosion at the
point where the stream impacts the disk (e.g. Livio, Soker \& 
Dgani 1986). Material expands
away from this point in all directions, and there is no
coherent stream flow over the disk. Similar
fractions of the incident gas stream flow inward in
either regime, but in the case of inefficient cooling
the impact leads to a more vertically extended
`halo' above and below the disk, and a bulge extending 
along the disk rim. The velocity of gas viewed along a
line-of-sight to the accreting star displays a much 
greater dispersion in the case of inefficient cooling, 
and has a mean velocity that has a smaller magnitude 
than if cooling is efficient and the stream overflows 
ballistically.

Aside from the strength of cooling, the most important
parameter in determining the degree of stream overflow
is likely to be the ratio of stream to disk scale heights
at the disk edge.
We have quantified this for the case of efficient cooling.
Assuming, as we have here, that both the stream and the 
disk have vertical density profiles that can be described as
gaussians, then $H_{\rm s} / H_{\rm d} \gtrsim 2$ 
is required for
significant stream overflow at ballistic velocities 
to occur (at lower ratios significant supersonic
inflow at lesser velocities may still be produced). 
We find that the numerical results can be reasonably
fit by a simple expression that assumes that the overflowing
mass fraction scales with the fraction of the stream
above the height where the disk and stream densities are
equal.

The simulations reported here have used very simple 
isothermal vertical disk profiles, and this is probably
the largest remaining uncertainty in the calculations.
A density profile that declines markedly less steeply
than the isothermal one would lead to stronger interaction
between the stream and the disk at several scale heights
above the midplane. The observations discussed above 
suggest that overflow can occur in CVs, but in 
extreme cases such a density profile could inhibit
overflow entirely. Even with weaker interactions,
the influence of the stream on a disk wind or on 
large-scale magnetic structures anchored to the disk surface
might extend the non-axisymmetric effects of the stream
to larger elevations above the disk surface than the
purely hydrodynamic effects studied in this paper.

\acknowledgments
We thank Steve Lubow, Norm Murray and Paula Szkody for 
helpful discussions, and the referee for many helpful 
suggestions that improved the presentation of this work.
P.J.A. thanks Space Telescope Science Institute for 
hospitality during the writing of this paper.
M.L. acknowledges support from
NASA Grant NAGW-2678.

\newpage

\begin{figure}[tb]
\caption{Isodensity surfaces from the isothermal (left column) and
        adiabatic (right column) simulations. The upper panels show 
        the top view, with the stream flowing left to right and the 
        disk flowing downwards; the lower panels show side views in which
        the disk material is moving out of the plane of the paper. 
        The isothermal calculation is rendered at a density surface 
        that is $10^{-3}$ of the central disk density. 
        The adiabatic run is rendered at a density surface 
        that is $10^{-2.75}$ of the central disk density. 
        {\em This figure available as a jpeg in the astro-ph version.
        Interactive viewing of the simulations at 
        {\tt http://www.cita.utoronto.ca/$^\sim$armitage/hydro{\underline{
        }}abs.html}}.}
\end{figure}        

\begin{figure}[tb]
\plotone{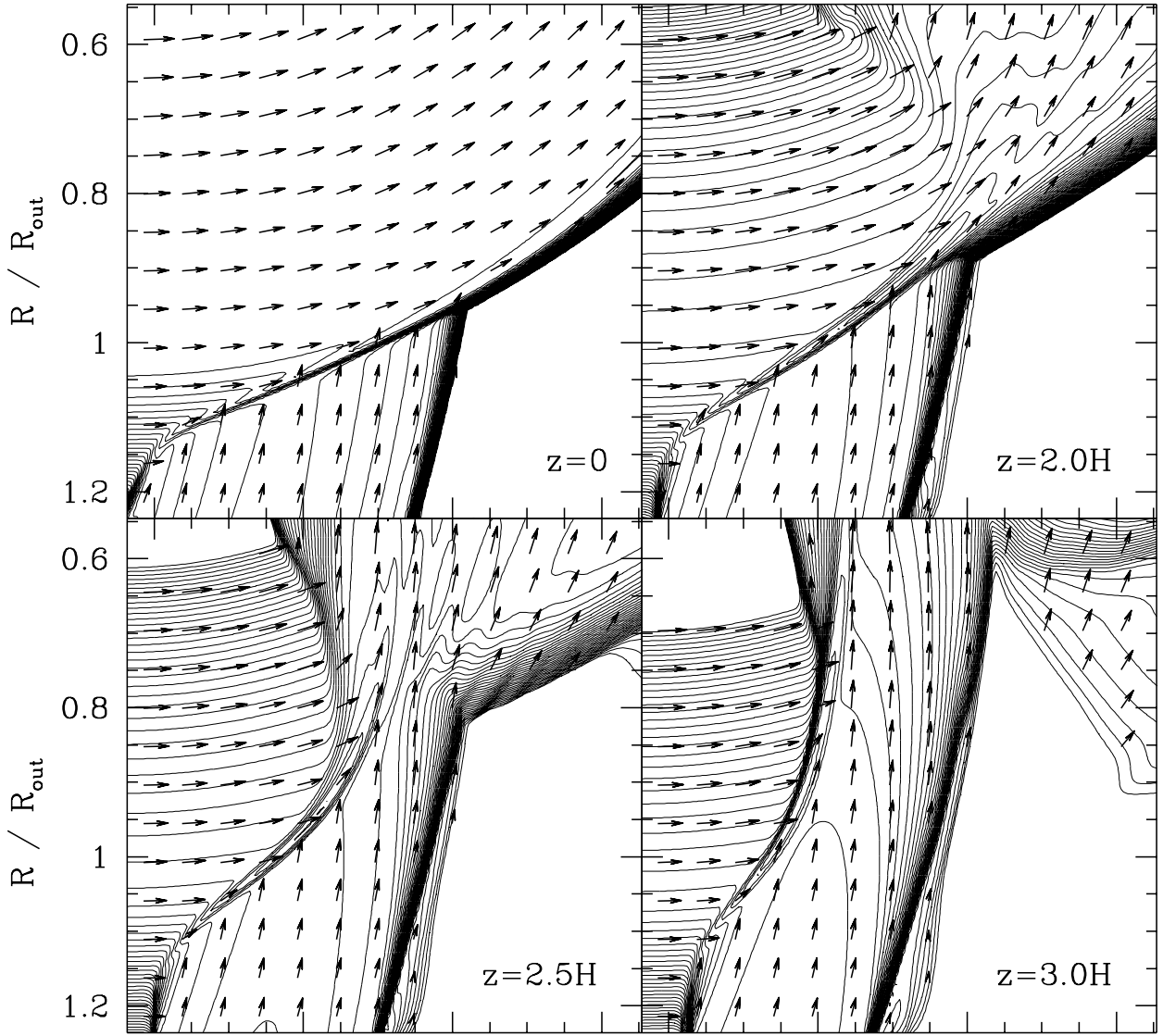}
\caption{Density contours in the $x-y$ plane for a calculation with 
	an isothermal equation of state. The computational grid was
	$176 \times 176 \times 55$ cells. $H_{\rm s} / H_{\rm d} = 2$.
        The panels depict slices
	in the disk midplane ($z=0$), and at 2, 2.5 and 3 {\em disk}
	scale heights above the plane. The stream enters from the
	bottom in each panel, the disk flow is left to right.
	Contour levels are at 
	$\Delta \log \rho = 0.3$, between $\log \rho = 0$ and 
	$\log \rho = -12$. Arrows depict the direction and 
	relative magnitude of the velocity field in each plane.
	The radius as a fraction of the disk radius $R_{\rm out}$
	is also shown.}
\end{figure}	

\begin{figure}[tb]
\plotone{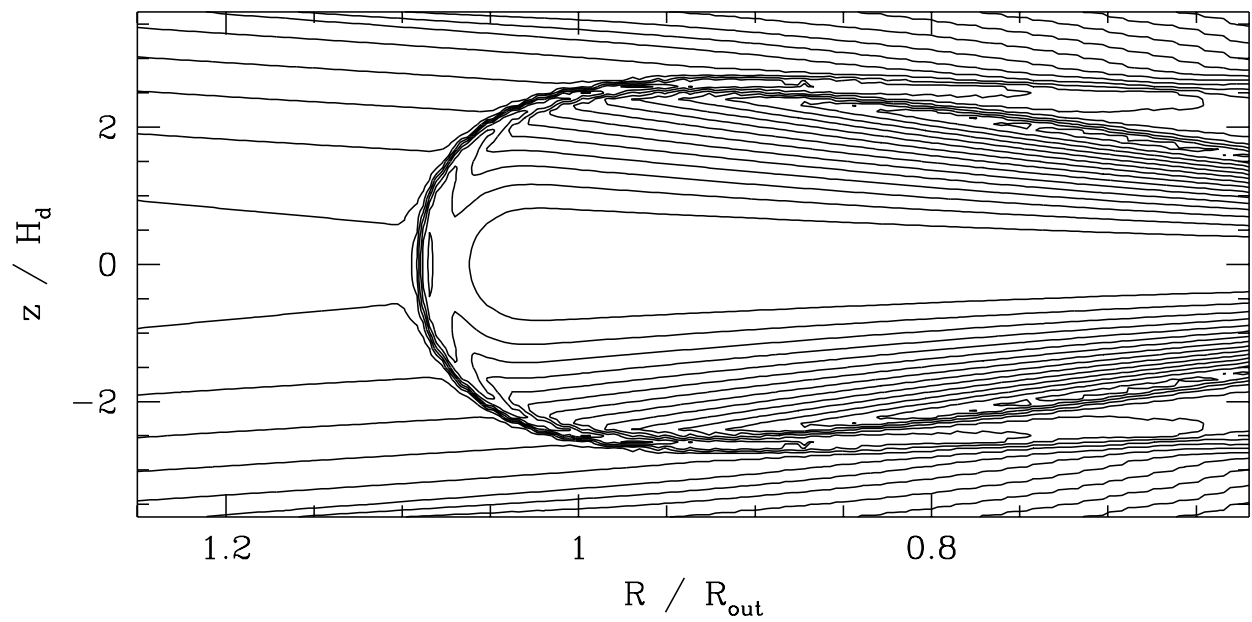}
\caption{Density contours in a vertical slice along the initial 
	stream flow direction for the isothermal calculation. Contour 
	levels are at $\Delta \log \rho = 0.3$.}
\end{figure}	
	
\begin{figure}[tb]
\plotone{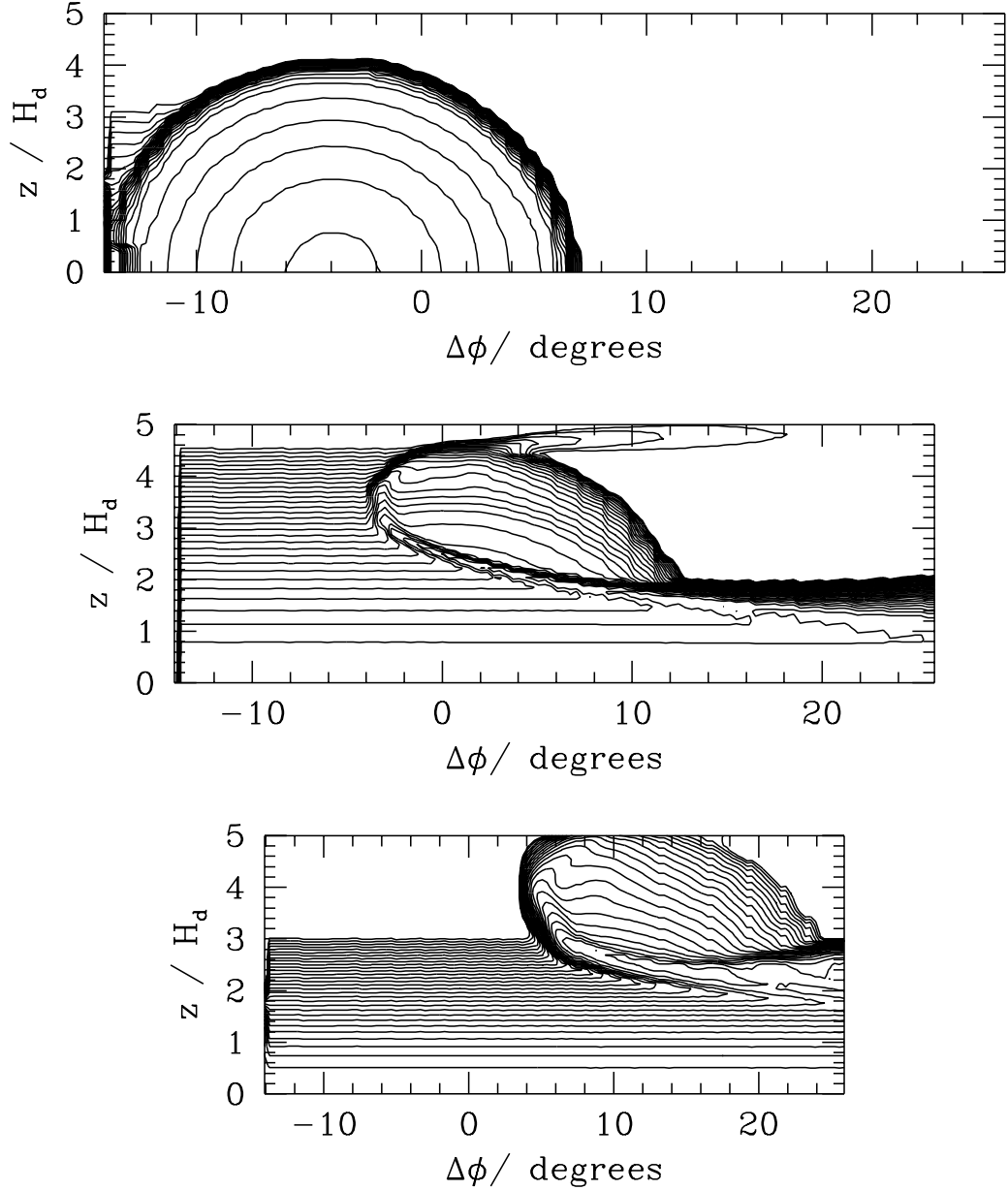}
\caption{Density contours on cylindrical surfaces at constant radius
	from the accreting object ($R \phi$,$z$ surfaces). From top
	downwards, the surfaces are at radii of $1.2 R_{\rm out}$,
	$1.0 R_{\rm out}$ and $0.75 R_{\rm out}$. Contour levels are at 
	$\Delta \log \rho = 0.3$, between $\log \rho = 0$ and 
	$\log \rho = -9$. The $x$-axis is labelled with the angular 
	distance relative to where the stream meets the disk edge -- in 
	this figure positive $\Delta \phi$ implies {\em downstream} 
	of the impact point.}
\end{figure}	

\begin{figure}[tb]
\plotone{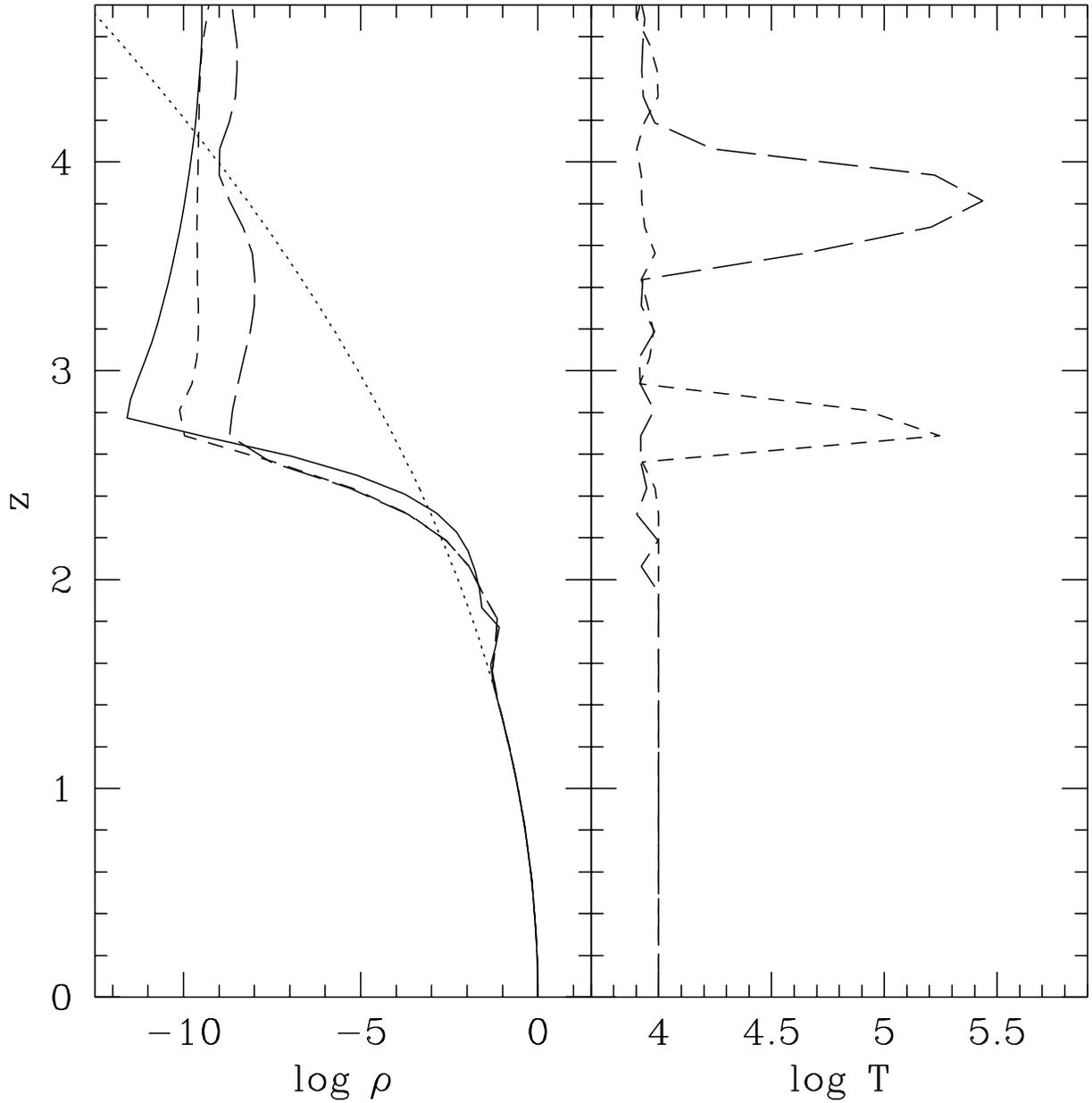}
\caption{Illustration of the influence of the cooling rate 
	on the vertical density and temperature profile downstream
	of the impact point on the disk rim (at $R = 0.9 R_{\rm out}$,
	and $20^\circ$ downstream of the stream impact position).
	The solid line shows the result for an isothermal 
	equation of state, the short dashed and long dashed lines
	for calculations including radiative cooling with central
	disk densities of $10^{14} {\rm cm}^{-3}$ and $10^{13} {\rm cm}^{-3}$
	respectively. The dotted line in the density profile 
	depicts the hydrostatic fall-off of density with height.}
\end{figure}	

\begin{figure}[tb]
\plotone{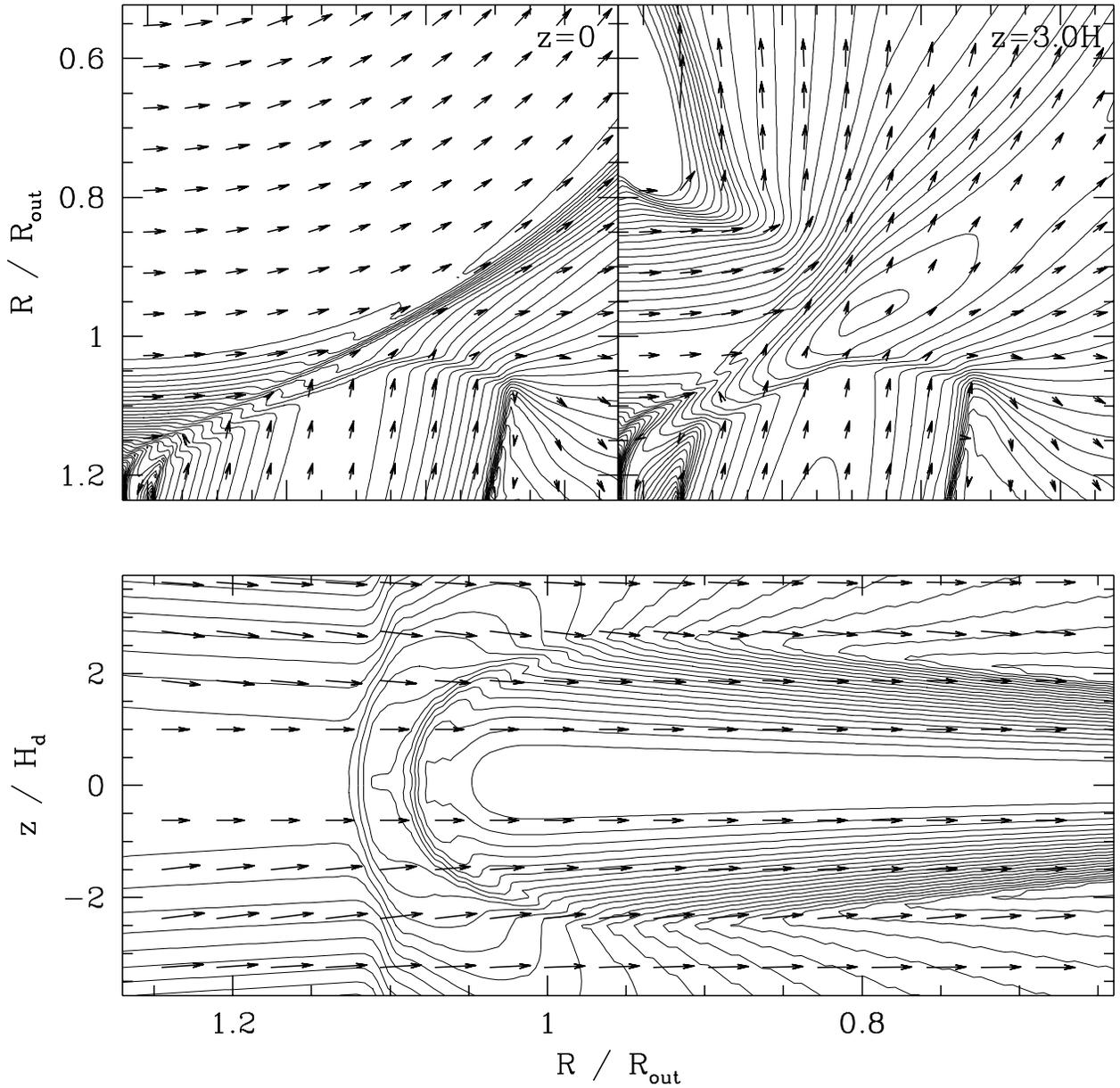}
\caption{Summary of results for an adiabatic equation of state,
	with initial conditions corresponding to the isothermal 
	calculation. The computational grid was
	$128 \times 128 \times 40$ cells. The upper panels
	show density contours in the $x-y$ plane at $z=0$ and
	at 3 disk scale heights, the lower panel a vertical
	slice along the initial stream flow direction. Contour 
	levels are at $\Delta \log \rho = 0.2$. Arrows depict 
	the direction and relative magnitude of the velocity 
	field in each plane. Note that the vertical slice 
	is not in the radial direction, this is why there are
	non-zero velocities even within the disk material.}
\end{figure}	

\begin{figure}[tb]
\caption{Results from SPH calculations of the entire disk
	with varying polytropic equations of state. Plotted 
	is a side view of the system with the stream entering
	from the left. Upper panel
	is for an isothermal equation of state, lower panel
	from a calculation with $\gamma = 1.1$.
	{\em Omitted for reasons of space. Armitage \& Livio (1996) 
	shows a similar figure}.}
\end{figure}	

\begin{figure}[tb]
\plotone{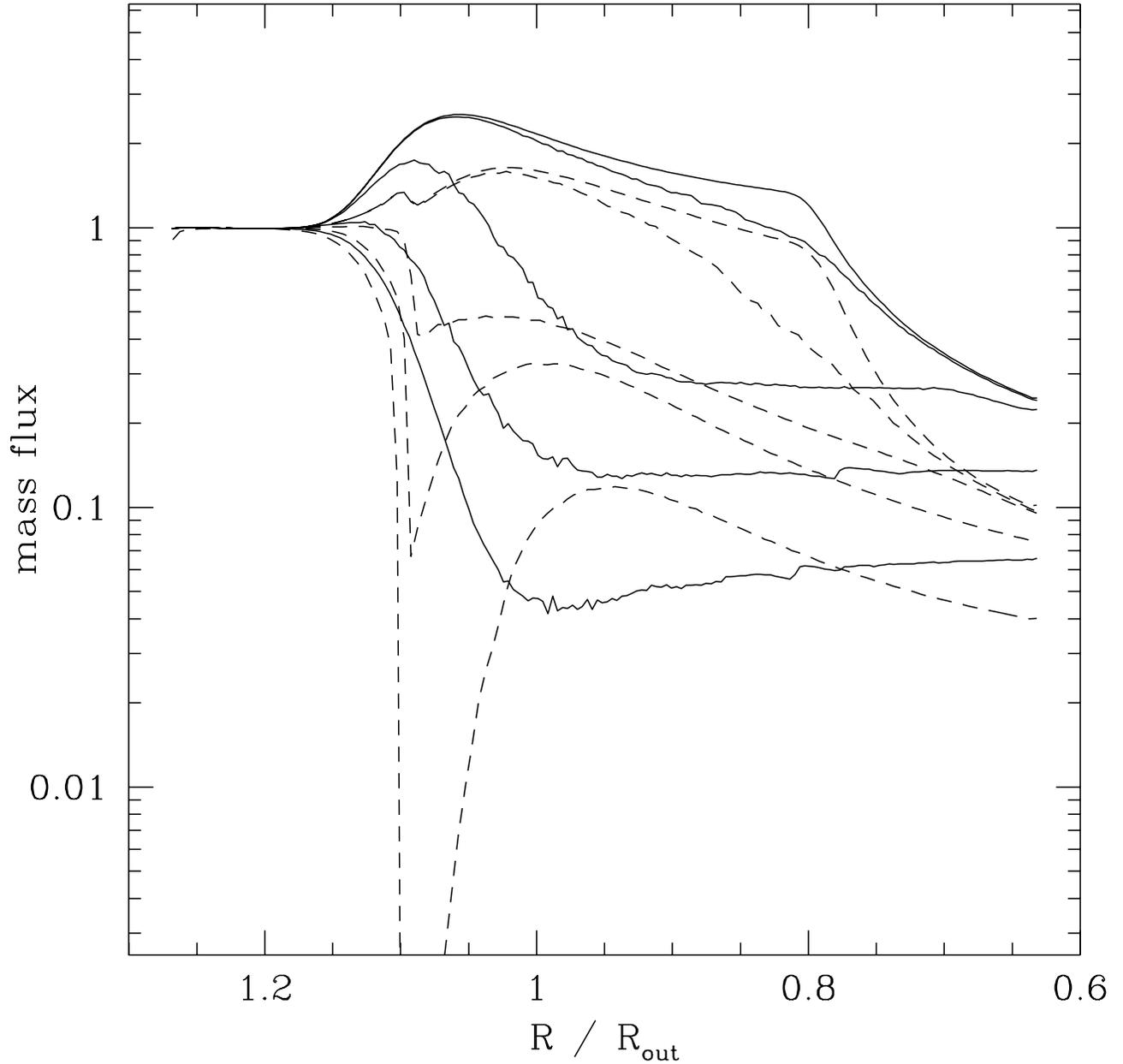}
\caption{Mass flux with inward radial velocity $v_R \ge v_{\rm cut}$ 
	through vertical slices at varying $y$ (labelled with the 
	fractional radius $R / R_{\rm out}$ measured along the initial
	stream flow direction). The solid lines show the results
	from an isothermal simulation, The dashed lines are for an adiabatic 
	calculation. Both simulations have $H_{\rm s} / H_{\rm d} = 2$.
	From top downwards, $v_{\rm cut} = 
	0$, $c_s$, $5 c_s$, $10 c_s$ and $20 c_s$.}
\end{figure}	

\begin{figure}[tb]
\plotone{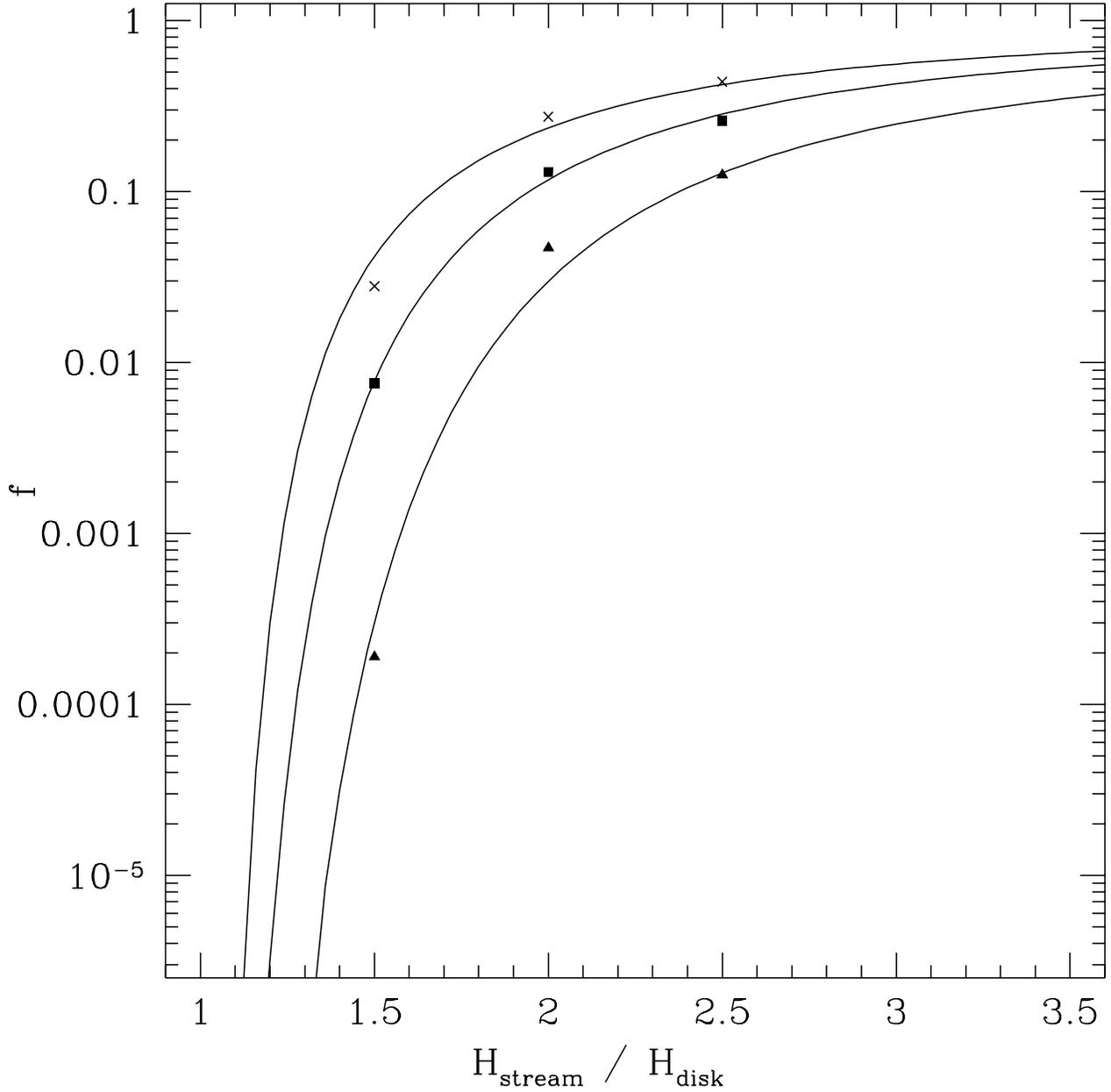}
\caption{Overflowing mass as a fraction $f$ of the stream mass
	flux, as a function of the ratio of scale heights of the 
	stream to the disk $H_{\rm stream} / H_{\rm disk}$. The
	symbols are from the numerical simulations; crosses for
	mass inflow at $v_R \ge 5 c_s$, squares for $v_R \ge 10 c_s$,
	and triangles for $v_R \ge 20 c_s$. The curves represent
	the scaling function described in the text, with 
	$\beta = $ 0.85, 1.1 and 1.5 respectively.}
\end{figure}	

\begin{figure}[tb]
\plotone{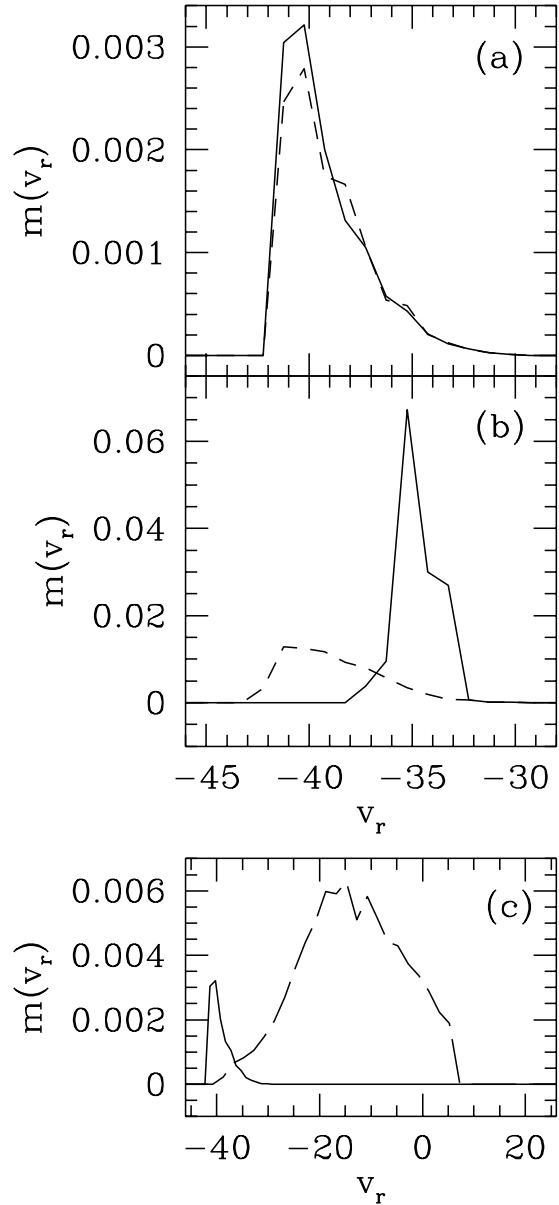}
\caption{The radial velocity distribution of gas along 
	lines of sight to the central accreting object. All
	panels are for a line of sight situated 18$^\circ$
	downstream of the impact point. (a) Elevation angle
	above the midplane $\theta = 12^\circ$. Solid line
	is the result from a simulation including cooling, 
	dashed line is the contribution from a ballistic
	stream with no disk interaction. (b) As in the 
	previous panel except with $\theta = 9^\circ$.
	(c) Comparison of an adiabatic (long dashed line)
	simulation with a simulation including efficient
	cooling (solid line), for $\theta = 12^\circ$ 
	(note different scale on the $x$ axis).}
\end{figure}	

\end{document}